\documentclass[pdflatex,sn-chicago]{sn-jnl}

\usepackage{graphicx}%
\usepackage{multirow}%
\usepackage{amsmath,amssymb,amsfonts}%
\usepackage{amsthm}%
\usepackage{mathrsfs}%
\usepackage[title]{appendix}%
\usepackage{xcolor}%
\usepackage{textcomp}%
\usepackage{manyfoot}%
\usepackage{booktabs}%
\usepackage{algorithm}%
\usepackage{algorithmicx}%
\usepackage{algpseudocode}%
\usepackage{listings}%

\usepackage{bm}

\newcommand{\Sb}{\bm{S}}
\newcommand{\Sigmab}{\bm{\Sigma}}

\newcommand{\Vb}{\bm{V}}

\newcommand{\Xb}{\bm{X}}

\theoremstyle{thmstyleone}%
\newtheorem{lemma}{Lemma}%
\theoremstyle{thmstyletwo}%
\newtheorem{remark}{Remark}%
\newtheorem{procedure}{Procedure}%

\theoremstyle{thmstylethree}%

\begin{document}

\title[Beyond Regularization: Inherently Sparse Principal Component Analysis]{Beyond Regularization: Inherently Sparse Principal Component Analysis}


\author[1,2]{\fnm{Jan O.} \sur{Bauer}}\email{j.bauer@vu.nl}

\affil[1]{\orgdiv{Department of Econometrics and Data Science}, \orgname{Vrije Universiteit Amsterdam}, \orgaddress{\street{De Boelelaan 21105}, \postcode{1081 HV}, \state{Amsterdam}, \country{The Netherlands}}}

\affil[2]{\orgname{Tinbergen Institute}, \orgaddress{\street{Gustav Mahlerplein 117}, \postcode{1082 MS}, \state{Amsterdam}, \country{The Netherlands}}}


\abstract{Sparse principal component analysis (sparse PCA) is a widely used technique for dimensionality reduction in multivariate analysis, addressing two key limitations of standard PCA. First, sparse PCA can be implemented in high-dimensional low sample size settings, such as genetic microarrays. Second, it improves interpretability as components are regularized to zero. However, over-regularization of sparse singular vectors can cause them to deviate greatly from the population singular vectors, potentially misrepresenting the data structure. Additionally, sparse singular vectors are often not orthogonal, resulting in shared information between components, which complicates the calculation of variance explained. To address these challenges, we propose a methodology for sparse PCA that reflects the inherent structure of the data matrix. Specifically, we identify uncorrelated submatrices of the data matrix, meaning that the covariance matrix exhibits a sparse block diagonal structure. Such sparse matrices commonly occur in high-dimensional settings. The singular vectors of such a data matrix are inherently sparse, which improves interpretability while capturing the underlying data structure. Furthermore, these singular vectors are orthogonal by construction, ensuring that they do not share information. We demonstrate the effectiveness of our method through simulations and provide real data applications. Supplementary materials for this article are available online.}




\maketitle

\section{Introduction \label{section intro}}

High-dimensional, low-sample size data (HDLSS) are becoming increasingly common in various areas of modern science, including genetic microarrays, medical imaging, text recognition, finance, and chemometrics. Principal component analysis (PCA) has long been a key tool for dimensionality reduction, particularly when data dimensionality is very high. By approximating the data with the first few principal components, PCA aims to visualize important structures and extract meaningful patterns. These components are computed using the (right) singular vectors of the data matrix.

However, PCA faces challenges in HDLSS settings. It is known to be inconsistent under such conditions (see, e.g., \citet{BS06}, \citet{NA08}, \citet{JL09}, \citet{JM09}, \citet{BJNP13}, and references therein). Another disadvantage of PCA is that each principal component (PC) is a linear combination of all the original variables, with the corresponding singular vectors (principal component loadings) usually having nonzero components. This lack of sparsity makes the interpretation of the results complicated.

To address the first drawback, assumptions where derived under which the singular value decomposition (SVD) remains consistent in HDLSS settings (see, e.g., \citet{JL09}, \citet{YA10}, \citet{BR13}, \citet{VL13}, \citet{WF17}, \citet{CHP20}, and references therein). These assumptions typically involve some form of sparsity in the singular vectors or the presence of spiked eigenvalues in the covariance matrix that lie outside the bulk spectrum. Furthermore, sparse PCA has been developed as a response to the interpretability challenges posed by traditional PCA. Sparse PCA introduces regularizations or constraints that shrink components of the singular vectors towards zero, thereby achieving sparsity (see, e.g., \citet{ZHT06}, \citet{SH08}, \citet{WTH09}, \citet{YMB14}, \citet{BHD20}, \citet{GWS20}, among others). Sparse PCA not only overcomes the challenge of interpretability, but is also suitable for HDLSS settings, making it a robust alternative to classical PCA in high-dimensional data analysis.

However, if the sparse singular vectors are over-regularized, they can deviate greatly from the population singular vectors, leading to overestimation or underestimation of the corresponding singular values and, consequently, the explained variance. Furthermore, these sparse singular vectors are often not orthogonal, leading to shared information between components. This lack of orthogonality complicates the calculation of the explained variance, as the contributions of the individual components must be disentangled.

In this work, we therefore propose a methodology to compute sparse singular vectors that reflect the inherent structure of the data matrix. While the assumption of sparsity in the first few singular vectors is not new (see, e.g, \citet{AW09}, \citet{MA13}, and \citet{CMW15}), our approach differs in important ways as we focus on sparse covariance matrices. Specifically, we consider data matrices that consist of uncorrelated submatrices, meaning that the covariance matrices exhibit block diagonal structures. This framework has recently been explored in different works (see, e.g., \citet{DG18}, \citet{BA24}, or \citet{SWB24}). Under such a structure, the right singular vectors of the data matrix, which align with the singular vectors (eigenvectors) of the covariance matrix, are inherently sparse. Our proposed method computes sparse PCs that naturally align with this structure, enhancing interpretability while preserving the representation of the underlying data. Additionally, these singular vectors are orthogonal by construction, which means they do not share information about the explained variance.

Another closely related approach to multivariate analysis is principal loading analysis \citep{BD21, BD23}. This methodology aims to identify disjoint subsets of variables in the data matrix that do not share information, and to select those subsets that explain most of the variance in the data. The concept traces back to \citet{MC84}, who introduced the similar concept of \emph{principal variables}, which are variables that capture as much information as possible, in response to the interpretability challenge in PCA, where each PCs is a combination of all variables. Since the work by \citet{MC84}, various approaches for detecting these principal variables have been developed (see, e.g., \citet{CCM04}, \citet{CW07}, \citet{BR14}, or \citet{CM24}). In principal loading analysis, detecting principal submatrices is typically achieved through hard thresholding of the singular vectors of the data matrix, a method that depends heavily on consistent estimation. In contrast, this paper proposes using sparse singular vectors to eliminate the need for hard thresholding.

\begin{figure*}[h]
\center
\includegraphics[width=.9\textwidth]{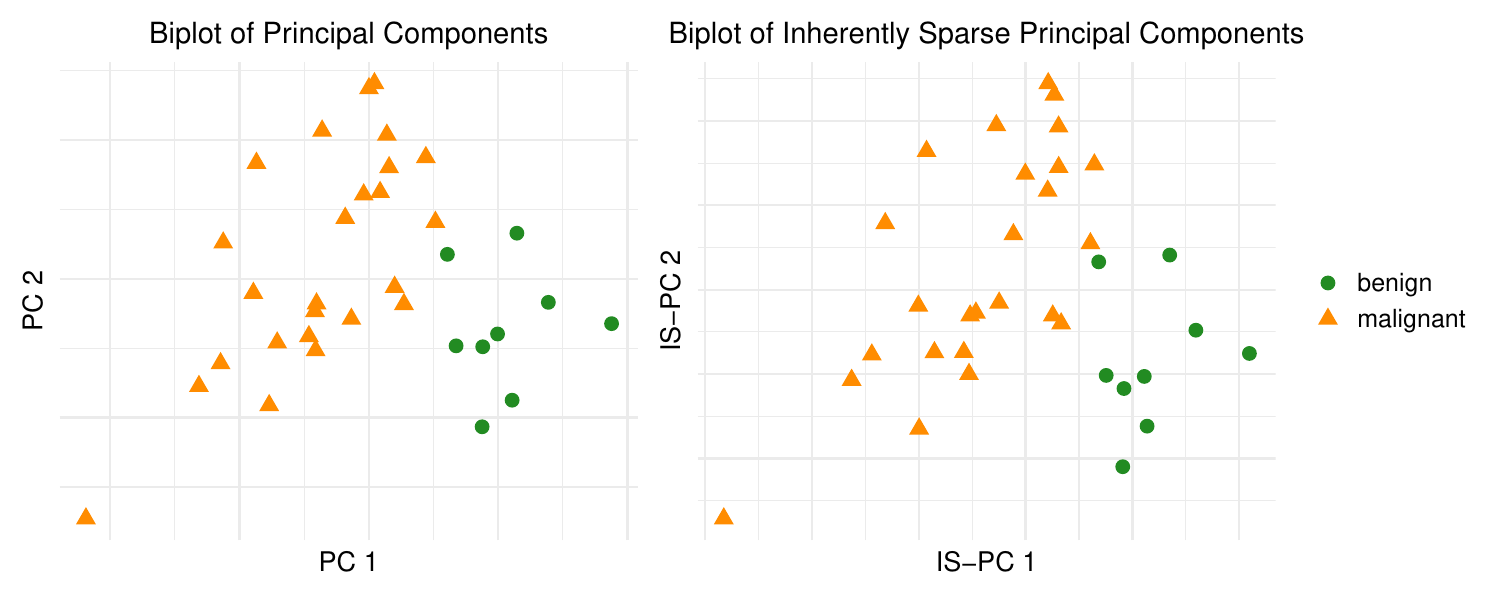}
  \caption{Biplots for a gene expression data matrix comprising $9$ benign (noncancerous) prostate tumors \emph{(green, $\bullet$)} and $25$ malignant (cancerous) observations \emph{(orange, $\blacktriangle$}). In the left plot, PCA is performed on the full $34 \times 12600$ data matrix which includes all gene expressions. In the right plot, PCA is performed on a $34 \times 219$ submatrix which thus contains only $219$ gene expressions.}
  \label{fig:ProstateIntroduction}
\end{figure*}

An illustration of the two methodologies proposed in this work is provided in Figure~\ref{fig:ProstateIntroduction}, which presents the first two PCs of a gene expression dataset containing $9$ benign (noncancerous) prostate tumor observations and $25$ malignant (cancerous) prostate tumor observations. In the left plot, PCA is performed on the full $34 \times 12600$ data matrix which includes all gene expressions. The resulting biplot captures differences between the two tumor types. In the right plot, we show a biplot of a $34 \times 219$ submatrix which thus contains only $219$ gene expressions. This submatrix captures $66.81\%$ of the total variance, seemingly containing a substantial amount of the variance within the data. Furthermore, this biplot distinguishes similarly between the two tumor types, indicating that the submatrix captures the same key information as the full data matrix. Effectively, the resulting PCs are sparse, as they highlight $219$ out of $12600$ genes. Computational details, data information, and further elaborations are provided in Section~\ref{s:RealData}.

The rest of this article is organized as follows. In Section~\ref{s:Methodology}, we introduce basic notations and definitions, and present the methodological details. We provide a simulation study and real data examples in Section~\ref{s:Results}. Section~\ref{s:Discussion} contains the discussion.

\section{Material and methods}
\label{s:Methodology}

\subsection{Setup}

Let $\bm{X} = (\bm{x}_1 , \ldots , \bm{x}_p) = ( \bm{X}_1 , \ldots, \bm{X}_b )$ be an independent and identically distributed $n \times p$ data matrix. $\bm{X}$ contains $p$ variables observed as $n \times 1$ column vectors $\bm{x}_1 , \ldots , \bm{x}_p $, and is partitioned into $b$ distinct submatrices $\bm{X}_i$, each of dimension $n \times p_i$ for $i \in \{ 1 , \ldots, b \}$ where $p = p_1 + \cdots + p_b$. Each $\bm{X}_i$ is organized such that $\bm{X}_1 = (\bm{x}_1 , \ldots , \bm{x}_{p_1})$ contains the first $p_1$ columns of the data matrix $\bm{X}$, $\bm{X}_2$ contains the next $p_2$ columns, and so forth. We can assume this convenient ordering as it can be obtained by column permutation. The SVD of the data matrix is expressed as:
\begin{equation*}
    \bm{X} = \bm{U} \bm{D} \bm{V}^T, \; \bm{U}^T\bm{U} = \bm{I}_n, \; \bm{V}^T\bm{V} = \bm{I}_p, \; d_1 \geq \cdots \geq d_r > 0 \, ,
\end{equation*}
where $r \leq \min(n,p)$ is the rank of $\bm{X}$, and $\bm{U} = (\bm{u}_1,\ldots,\bm{u}_n)$ and $\bm{V} = (\bm{v}_1,\ldots,\bm{v}_p)$ are the left and right singular vectors respectively. Without loss of generality, assume that the columns of $\bm{X}$ are mean zero. This implies that the right singular vectors of the data matrix align with the singular vectors (eigenvectors) of the sample covariance matrix $\bm{S} = n^{-1} \bm{X}^T\bm{X} $. The SVD of $\bm{S}$ is given by
\begin{equation*}
    \bm{S} = \bm{V} \bm{L} \bm{V}^T, \; \bm{V}^T \bm{V} = \bm{I}_p, \; l_1 \geq \cdots \geq l_r > 0 \, .
\end{equation*}
Here, $\bm{L}$ contains the singular values (eigenvalues) $l_j$ for $j \in \{1, \ldots, p\}$ of $\bm{S}$ on the diagonal.

We assume that the unobserved population covariance matrix $\Sigmab$ follows a block diagonal structure:
\begin{equation}\label{eq:SigmaBlockDiagonal}
 \Sigmab \equiv   \begin{pmatrix}
    \Sigmab_1 & \bm{0} & \bm{0} \\
    \bm{0} & \ddots & \bm{0}\\
    \bm{0} & \bm{0} & \Sigmab_b
    \end{pmatrix} \, ,
\end{equation}
where $\Sigmab_i$ is the population covariance matrix corresponding to the submatrix $\Xb_i$ for $i\in\{1,\ldots,b\}$, and the population covariance between these submatrices is $\bm{0}$. In practice, we have only access to the sample covariance matrix $\Sb$ of the observed data.

\subsection{Principal submatrices}

Assume for now that $\bm{S} = \bm{\Sigma}$ such that the population covariance matrix is perfectly observed by its sample counterpart.  It holds that the right singular vectors of $\bm{X}$ mirror the block diagonal structure of the covariance matrix:
\begin{lemma}
\label{l:VectorStructure}
$\Sb$ is a block diagonal matrix if and only if its singular vectors (the right singular vectors of the submatrix $\Xb$) exhibit the structure
\begin{equation}\label{eq:VectorBlockStructure}
    \Vb \bm{P}_\pi =  \begin{pmatrix}
    \Vb_1 & \bm{0} & \bm{0} \\
    \bm{0} & \ddots & \bm{0}\\
    \bm{0} & \bm{0} & \Vb_b
    \end{pmatrix} \, ,
\end{equation}
where $\Vb_i$ are the singular vectors of $\Sb_i$ (the right singular vectors of the submatrix $\Xb_i$) for $i \in \{1,\ldots,b\}$, and $\bm{P}_\pi$ is a permutation matrix leading to a block diagonal structure.
\end{lemma}
Each submatrix captures the variance in different orthogonal directions, meaning that they do not share any linear information about the variance because they are uncorrelated: $\bm{X}_i^T \bm{X}_j = \bm{0}$ for $i\neq j$. Building upon this, \citet{BD21} introduced the concept of \emph{principal loading analysis} which computes the explained variance for each submatrix and retains only the variables associated with submatrices that contribute substantially to the total explained variance. These submatrices are referred to as \emph{principal submatrices} in this work.

\begin{remark}[Principal loading analysis from \citet{BD21}]\label{re:PLA}
The explained variance of the variables contained in submatrix $\bm{X}_i$ in comparison to the overall explained variance equals
\begin{equation}\label{eq:PopExpVar}
 \sum_{k=1}^{p_i} l_{i,k} \,/\, \sum_{j = 1}^p l_j  \,.
\end{equation}
Here, $l_{i,1} , \ldots, l_{i, p_i}$ are the singular values (eigenvalues) of the covariance matrix $\bm{S}_i$ corresponding to the submatrix $\bm{X}_i$.
\end{remark}

Recall that we previously assumed $\bm{S} = \bm{\Sigma}$. In practice, this equality rarely holds and the sample covariance matrix is rather a perturbed transformation of the population covariance matrix
\begin{equation*}
\Sb \equiv \Sigmab + \bm{E}  \, .
\end{equation*}
Therefore, identifying the block diagonal structure of both the covariance matrix and the eigenvectors is not trivial because it is masked by the noise matrix $\bm{E}$.

Initially, \citet{BD21} proposed using hard thresholding on the eigenvectors to uncover the underlying block diagonal structure. However, this approach imposes several challenges: First, the threshold value is based only on simulation studies, making it difficult to determine an appropriate value in applications. Second, the method relies on precise estimation of the singular vectors to prevent misspecification after hard thresholding. Third, all $p$ eigenvectors of the covariance matrix must be evaluated, as some components in the leading eigenvectors might be set equal to zero.

\citet{BA24} recently proposed a method to identify the population block diagonal covariance structure using singular vectors (BD-SVD). This approach overcomes the challenges mentioned above and can thus be used to detect all submatrices. The key idea of BD-SVD is to use sparse approximations of the right singular vectors $\bm{V}$ of the data matrix $\bm{X}$. For completeness, we recap BD-SVD in the supplementary data.

Additionally, we recalled in Remark~\ref{re:PLA} that the explained variance for each submatrix $\bm{X}_i$ is given by the eigenvalues of $\bm{S}_i$ as in \eqref{eq:PopExpVar} if the population covariance matrix is perfectly observed, which means if $\bm{S} = \bm{\Sigma}$. In practice, when $\bm{S} = \bm{\Sigma} + \bm{E}$, these eigenvalues are perturbed by $\bm{E}$.

\begin{remark}[Weyl's inequality]\label{re:Weyl}
    Let $\lambda_1 , \ldots, \lambda_p$ be the eigenvalues of the population covariance matrix $\bm{\Sigma}$. It holds that
    \begin{equation*}
        | \lambda_j - l_j |   \leq \Vert \bm{E} \Vert_2 \,,
    \end{equation*}
    for $j \in \{1 , \ldots, p \}$. Here, $\Vert \cdot \Vert_2$ is the spectral norm.
\end{remark}

If $p<n$, it is known that the sample eigenvalues are consistent estimators for the population eigenvalues. For the $p\gg n$ (and $p>n$) case, the first population eigenvalues can be consistently estimated under the assumption of a spiked or power spiked model (see, e.g., \citet{JO01}, \citet{NA08}, \citet{YA13}, and references therein).

However, we can also compute the explained variance of each submatrix without the eigenvalues. Recall from \eqref{eq:PopExpVar} that we require the sum of the eigenvalues of the covariance matrix $\bm{S}_i$ of the submatrix $\bm{X}_i$ alongside the sum of all eigenvalues of $\bm{S}$. Since the sum of eigenvalues of any symmetric matrix equals the sum of its main diagonal elements (i.e., its trace), we can replace the sum of eigenvalues by the sum of the variables' variances. The full procedure for principal loading analysis is summarized in what follows.

\begin{procedure}\label{pr:SPLA} Principal loading analysis using BD-SVD for an $n \times p$ data matrix $\bm{X} $ with column mean zero:
\begin{enumerate}
\item\label{Step1} Identify the block diagonal structure of the unobserved population covariance matrix $\bm{\Sigma}$ as in \eqref{eq:SigmaBlockDiagonal}, thus the population covariance matrices $\bm{\Sigma}_1 , \ldots, \bm{\Sigma}_b$, using BD-SVD to effectively decompose $\bm{X}$ into its submatrices $\bm{X}_1 , \ldots, \bm{X}_b$.

\item\label{Step2} Compute the explained variance for each submatrix $\bm{X}_i$ as
\begin{equation*}
    \sum_{k=1}^{p_i} s_{i,h} / \sum_{j=1}^p s_{j} \,,
\end{equation*}
where $s_{j} = {\rm Var}(\bm{x}_j)$ for $j \in \{1 , \ldots, p\}$ are the main diagonal elements of $\bm{S}$ (the variances of all variables in $\bm{X}$), and $s_{i,h}$ for $k \in \{1 , \ldots, p_i\}$ are the main diagonal elements of the covariance matrix $\bm{S}_i$ of $\bm{X}_i$ (the variances of the variables contained in $\bm{X}_i$) respectively.
\end{enumerate}
\end{procedure}
Consequently, using this approach the estimation of the covariance matrix for principal loading analysis becomes obsolete since identifying the block structure and computing explained variances of the blocks can be achieved using the data matrix only. This is advantageous in HDLSS settings where estimating the covariance matrix is challenging.

\subsection{Inherently sparse principal component analysis}

PCA is a widely used technique for dimensionality reduction and interpretation of data matrices \citep{JWTT21}. It computes PCs that capture most of the explained variance by transforming the data matrix using its singular vectors. The variance of each of these PCs $\bm{X} \bm{v}_j$ for $j \in \{1, 2, \ldots \}$ equals the eigenvalue $l_j$.

However, interpreting the principal components can be challenging, as they are linear combinations of all variables. Sparse PCA addresses this issue by setting components of the singular vectors to zero, improving interpretability at the cost of a lower explained variance. Sparse PCA also enables computation settings where $p\gg n$. Typically, calculation is achieved through a regularization term as in (A2) in the online appendix.

Notably, if the sparse singular vectors are over-regularized and therefore fail to align with the inherent structure of the underlying data, they can deviate greatly from the population singular vectors. In this case, we can also expect incorrectly computed singular values. To address this issue, we propose constructing sparse PCs that mirror the inherently sparse structure of the data. However, existing sparse PCA methods are not specifically designed for this purpose, as \citet{BA24} demonstrated that such inherently sparse structures are not effectively detected.

We recall from \eqref{eq:VectorBlockStructure} in Lemma~\ref{l:VectorStructure} that the eigenvectors of a block diagonal matrix are inherently sparse. Consequently, if we detect an underlying block diagonal structure of the population covariance matrix, we can construct sparse singular vectors accordingly that mirror this inherently sparse structure.

\begin{procedure}\label{pr:SPCA} Inherently sparse principal component analysis (IS-PCA) for an $n \times p$ data matrix $\bm{X} $ with column mean zero:
\begin{enumerate}
\item Identify the data matrix $\bm{X} = (\bm{X}_1 , \ldots, \bm{X}_b)$ reflecting the block diagonal covariance structure using BD-SVD.

\item\label{SCPA:Step2} Compute the SVD $\bm{X}_i = \bm{U}_i \bm{D}_i \bm{V}_i^T$ for each submatrix $\bm{X}_i$ and pad the right singular vectors with zeros to construct the sparse PC loadings $\tilde{\bm{V}}$ of $\bm{X}$ as
\begin{equation}\label{eq:SPCA}
\tilde{\bm{V}} \tilde{\bm{P}}_\pi =  \begin{pmatrix}
     \bm{V}_1 & \bm{0} & \bm{0} \\
    \bm{0} & \ddots & \bm{0}\\
    \bm{0} & \bm{0} &  \bm{V}_b
    \end{pmatrix}\,,
\end{equation} 
which mirror the inherently sparse structure. Here, $\tilde{\bm{P}}_\pi$ is a permutation matrix leading to a block diagonal structure. 
\end{enumerate}
\textbf{Output:} (Inherently) sparse PCs $\bm{Z} = \bm{X}\tilde{\bm{V}}$ with loadings $\tilde{\bm{V}}$.
\end{procedure}

First, we address a notational nuance. Ignoring sample noise, the inherently sparse singular vectors $\tilde{\bm{V}}$ in \eqref{eq:SPCA} are equivalent to the singular vectors $\bm{V}$ of $\bm{X}$ in \eqref{eq:VectorBlockStructure}. However, in practical applications, this equivalence is not straightforward. The singular vectors $\bm{V}$ are computed on the full $n \times p$ data matrix, whereas the inherently sparse singular vectors $\tilde{\bm{V}}$ are computed on $n \times p_i$ submatrices, where $i \in \{1, \ldots , b \}$, with $p_i \leq p$ and $p_i = p$ if and only if $b = 1$. Intuitively, the computed inherently sparse singular vectors $\tilde{\bm{V}}$ are expected to differ from $\bm{V}$, as we demonstrate in the simulation study in Section~\ref{s:SimulationStudy}.

When constructing sparse PCs using regularization approaches, the resulting PC loadings (singular vectors) are not necessarily orthogonal which complicates computing the explained variance. The loadings in \eqref{eq:SPCA}, however, are orthogonal which means that the corresponding explained variance can calculated be using the eigenvalues as in PCA.

\begin{remark}\label{re:OrthAndExpVar}
The sparse PC loadings (singular vectors) $\tilde{\bm{V}} $ obtained in \eqref{eq:SPCA} using Procedure~\ref{pr:SPCA} are orthonormal, and the explained variance for each sparse PC contained in $\bm{Z} = \bm{X} \tilde{\bm{V}} $ equals its corresponding eigenvalue of $\bm{S}_1 , \ldots, \bm{S}_b$.
\end{remark}

When $p_i \gg n$, using the classic SVD yields inconsistent estimators of the underlying singular vectors and singular values as discussed in the introduction of this work. We can overcome this issue by using approaches derived for HDLSS settings. In this work, we use the cross data matrix approach (CDM) proposed by \citet{YA10} as it appears robust for different scenarios \citep{YA13, WHC20, WH22}. The resulting singular vectors are consistent in a sense that the angle between the estimated and the population singular vectors converges to zero in probability. While their approach yields estimated singular vectors that are orthonormal in the limit, orthonormality holds only approximately in the sample case. However, this means that the singular values are an approximation for the explained variance in the HDLSS setting. In the supplementary data, we elaborately discuss the computation of the singular vectors alongside their orthogonality.

\section{Results}\label{s:Results}

\subsection{Preamble}

All computational results were obtained using the statistical software \texttt{R} 4.1.3 \citep{RSoftware} on a PC running macOS Sonoma 14.6.1. with $8$ GB of RAM. Computational details together with code to replicate the results are available in the supplementary data.

\subsection{Simulation study}\label{s:SimulationStudy}

\subsubsection{Simulation study design}

In this section, we highlight the advantage of IS-PCA in a simulation study. We simulate a data matrix with $n = 100$ observations and $p = 10000$ variables, drawn 100 times from a $p$-multivariate normal distribution $N(\bm{0}, \bm{\Sigma})$. The covariance matrix $\bm{\Sigma} = ( \bm{\Sigma}_1 , \ldots , \bm{\Sigma}_b)$ has a block diagonal structure with $b \in \{10, 20, 40, 50\}$ blocks of equal sizes. Each block $\bm{\Sigma}_i = (1 - \omega_i) \bm{I}_{p_i} + 2 \omega_i \bm{1}_{p_i} \bm{1}_{p_i}^T$ exhibits a compound symmetric covariance structure with a single spike, where $\omega_i$ is uniformly $U(0.1, 0.3)$ distributed and $\bm{1}_{p_i}$ is a vector of ones.

We evaluate five approaches for estimating the singular vectors and eigenvalues in this simulation study:

\begin{enumerate}
    \item[1.] \emph{CDM}: The CDM method as proposed by \citet{YA10}.
    \item[2.] \emph{Oracle Block IS-PCA}: IS-PCA as described in Procedure~\ref{pr:SPCA}, with known block structure. SVD is computed using CDM.
    \item[3.] \emph{False Negative IS-PCA}: IS-PCA as described in Procedure~\ref{pr:SPCA} but with an incorrect number of only $b/2$ blocks. This corresponds to a ``false negative'' scenario where blocks are not further split to identify the true structure. SVD is computed using CDM.
    \item[4.] \emph{PMD IS-PCA}: IS-PCA as described in Procedure~\ref{pr:SPCA}, but the number of blocks is identified using the sparse PCA (penalized matrix decomposition, PMD) method by \citet{WTH09}. SVD is computed using CDM.
    \item[5.] \emph{PMD}: The sparse PCA method as proposed by \citet{WTH09}.
\end{enumerate}

For the first four approaches, SVD is computed using the CDM method to ensure comparability. The key difference among these approaches lies in the number of blocks considered for IS-PCA: No blocks (\emph{CDM}), the true number of blocks (\emph{OB IS-PCA}), too few blocks (\emph{FN IS-PCA}), and blocks identified by sparse PCA (\emph{PMD IS-PCA}). As shown by \citet{BA24}, sparse PCA is not designed for block detection and often incorrectly continues to split blocks. This makes PMD IS-PCA a ``false positive'' version of IS-PCA.

For completeness, we also include a fifth approach, in which the SVD is directly calculated by PMD rather than by CDM. Similar to \citet{BA24}, in approaches four and five, only the first singular vector with corresponding eigenvalue is considered, as these alone already show poor results for the purpose of block detection.

To evaluate the accuracy of the estimated (sparse) singular vectors, we use the absolute cosine similarity, calculated as $| \bm{v}^T \tilde{\bm{v}}|$. This metric measures the alignment of two vectors, with values ranging from $0$ (orthogonal) to $1$ (same direction). For eigenvalues, we assess accuracy using the ratio $\tilde{l}/l$. A ratio greater than 1 indicates overestimation, while a ratio less than 1 indicates underestimation.

\subsubsection{Simulation study results}

\begin{figure*}[h]
\center
\includegraphics[width=.9\textwidth]{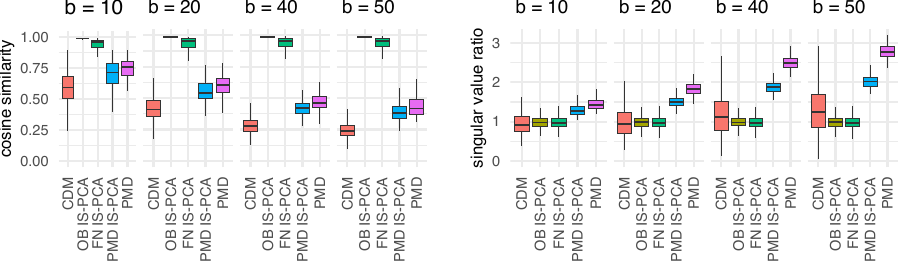}
  \caption{Simulation results for CDM \emph{(red)}, OB IS-PCA \emph{(yellow)}, FN IS-PCA \emph{(green)}, PMD IS-PCA \emph{(blue)}, and PMD \emph{(pink)} for the cosine similarity $| \bm{v}^T \tilde{\bm{v}}|$ \emph{(left plot)} and singular value ratio $\tilde{l}/l$ \emph{(right plot)}.}
  \label{fig:Simulation}
\end{figure*}

Simulation results are provided in Figure~\ref{fig:Simulation}. OB IS-PCA performs best, as PCA is computed on each underlying submatrix $\bm{X}_i$, effectively reducing the HDLSS ratio from $p/n$ to $p_i / n$, which results in better estimation accuracy. FN IS-PCA also demonstrates better performance compared to CDM, which does not utilize any block information. In contrast, PMD IS-PCA, the ``false positive'' approach based on sparse PCA, performs substantially worse. Notably, for this approach, the eigenvalues are strongly overestimated. This occurs because sparse PCA aims to capture as much explained variance as possible, leading it to select variables from multiple blocks to maximize variance across all blocks.

These findings underscore the importance of correctly identifying the underlying block diagonal structure for IS-PCA. Notably, in cases of false negative block detection, where not all blocks are split into their true structure, estimation of the SVD improves compared to CDM. In contrast, false positive block detection, where blocks are split incorrectly, reduces the performance of IS-PCA. In practice, opting for a less stringent penalization $a_{np}$ in the HBIC for BD-SVD (see (A4) in the online appendix) to avoid false positive block splits is therefore recommendable.

\subsection{Real data applications}\label{s:RealData}

\subsubsection{Prostate tumors}
This section analyzes a gene expression dataset of prostate tumors. The data set contains $p = 12600$ gene expressions from $n = 35$ tumors, of which $9$ are benign (noncancerous) and 25 are malignant (cancerous). We compute sparse PCA using IS-PCA alongside PCA, both by CDM.

\begin{figure*}[h]
\center
\includegraphics[width=.9\textwidth]{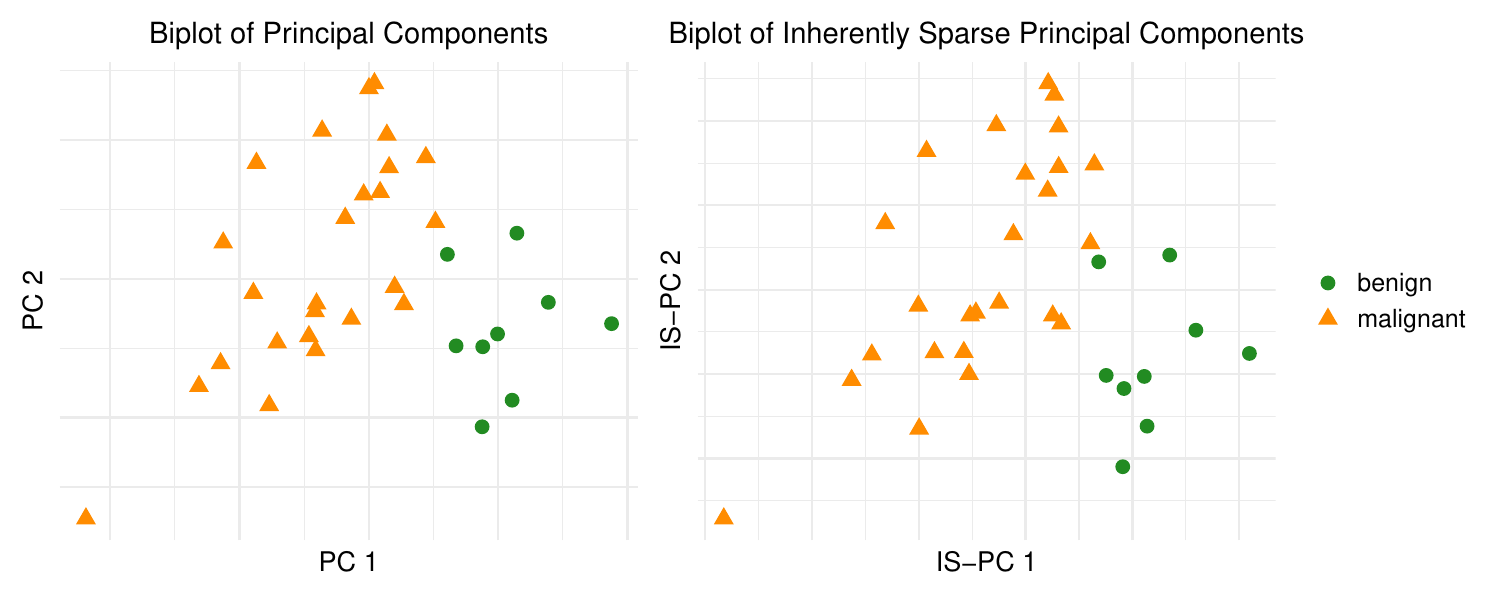}
  \caption{Biplots of the first PCs for the prostate tumors gene expression dataset, which contains $9$ benign (noncancerous) prostate tumors \emph{(green, $\bullet$)} and $25$ malignant (cancerous) observations \emph{(orange, $\blacktriangle$)}. The left plot shows PCA performed on the full $34 \times 12600$ data matrix which encompasses all gene expressions. The right plot provides the PCs from IS-PCA, consisting of sparse singular vectors with only $219$ nonzero components out of $12600$ meaning that PCA was performed on a $34 \times 219$ submatrix.}
  \label{fig:ProstateBiplot1}
\end{figure*}

We identify $b = 572$ submatrices, with one submatrix comprising $219$ variables that captures $66.81\%$ of the total variance, compared to an expected average contribution of $1.74\%$ ($219/12600 \approx 0.0174$). The first two (inherently) sparse PCs correspond to this principal submatrix. Further, the correlation between the first sparse singular vector and its non-sparse counterpart is $0.93$, while the second singular vectors have a correlation of $0.90$. Therefore, we expect that the resulting PCs obtained by IS-PCA and PCA contain similar information. Specifically, since the first two sparse singular vectors consist of only $219$ nonzero components out of $12600$, IS-PCA appears to capture the genes primarily responsible for the variance in the direction of the first two non-sparse singular vectors. Figure~\ref{fig:ProstateBiplot1} shows a biplot of the first two PCs, which is the identical figure as in Figure~\ref{fig:ProstateIntroduction} from the introduction.

\begin{figure*}[h]
\center
\includegraphics[width=.9\textwidth]{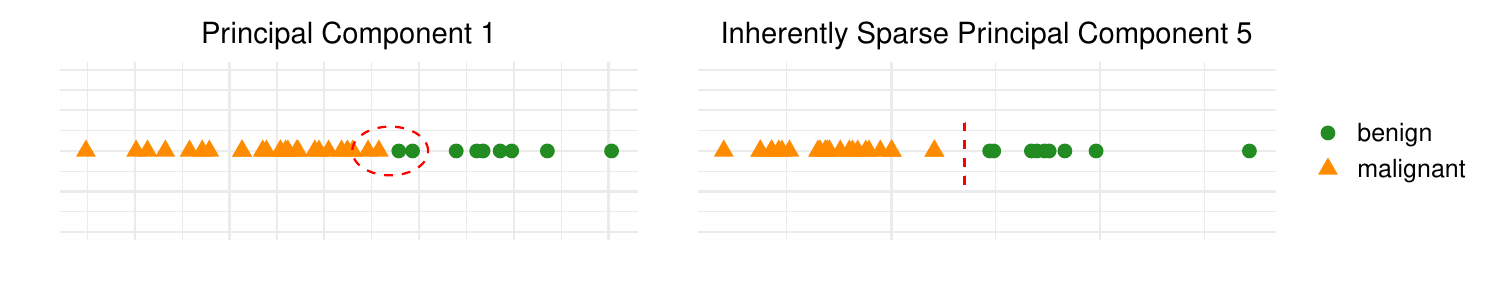}
  \caption{The first PC \emph{(left)} and fifth sparse PC \emph{(right)} for the prostate tumors gene expression data set, which contains $9$ benign (noncancerous) prostate tumors \emph{(green, $\bullet$)} and $25$ malignant (cancerous) observations \emph{(orange, $\blacktriangle$)}. The first PC does not to clearly separate the two tumor types, with some observations clustering together \emph{(red dashed circle)}. In contrast, the gene expressions in the submatrix underlying the fifth sparse PC seems to capture differences between the tumor types, visually distinguishing them \emph{(red dashed line)}.}
  \label{fig:ProstatePC5}
\end{figure*}

In both biplots in Figure~\ref{fig:ProstateBiplot1}, the benign tumors are mapped to the bottom/center-right. Consequently, neither the first PC (\emph{$x$-axis, left-right}) nor the second PC (\emph{$y$-axis, top-bottom}) solely distinguishes the two tumor types in a one-dimensional direction. Notably, the fifth PC in IS-PCA captures this information (Figure~\ref{fig:ProstatePC5}). This sparse singular vector has $223$ nonzero components and shows weak correlations with all non-sparse singular vectors, with the largest correlations being $0.23$ and $0.16$ for the first two singular vectors. Thus, it seems that the submatrix underlying the fifth sparse singular vectors extracts $223$ genes necessary to distinguish between benign and malignant tumors.

\subsubsection{Lymphoma malignancies}
In this section, we analyze a gene expression data set of lymphoid malignancies. The data set contains $p = 4026$ gene expressions from $n = 62$ samples: $42$ of diffuse large B cell lymphoma (DLBCL), $9$ of follicular lymphoma (FL), and $11$ of B cell chronic lymphocytic leukemia (CLL). We compute sparse PCA using IS-PCA alongside PCA, both by CDM.

\begin{figure*}[h]
\center
\includegraphics[width=.9\textwidth]{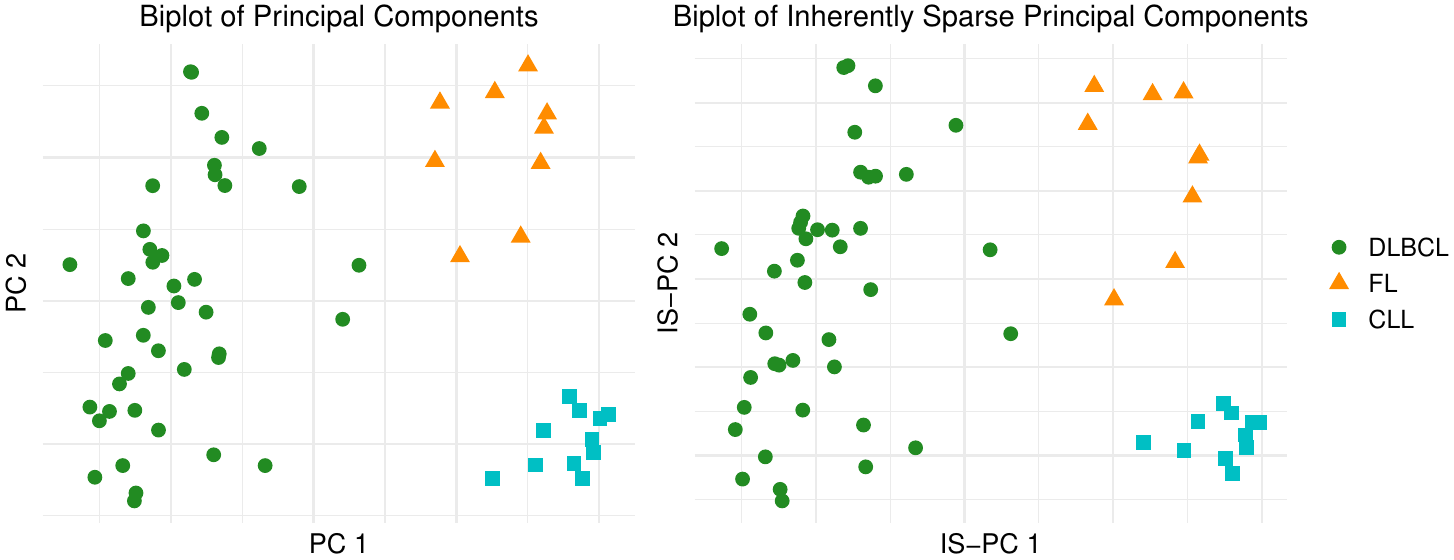}
  \caption{Biplots of the first PCs for the lymphoma gene expression data set, which contains $42$ diffuse large B cell lymphoma (DLBCL) observations \emph{(green, $\bullet$)}, $9$ follicular lymphoma (FL) observations \emph{(orange, $\blacktriangle$)}, and $11$ chronic lymphocytic leukemia (CLL) observations \emph{(blue, $\blacksquare$)}. The left plot shows PCA performed on the full $62 \times 4026$ data matrix which encompasses all gene expressions. The right plot provides the PCs from IS-PCA, consisting of sparse singular vectors with only $1901$ nonzero components out of $4026$ meaning that PCA was performed on a $62 \times 1901$ submatrix.}
  \label{fig:LymphomanaBiplot1}
\end{figure*}

The correlation between the first sparse singular vector and its non-sparse counterpart is $0.97$, while the second singular vectors have a correlation of $0.77$, suggesting that the resulting PCs capture similar information. However, the first two sparse singular vectors contain $1901$ nonzero components, effectively selecting genes that contribute the most to the information contained in the direction of the first two non-sparse singular vectors. A biplot of the first two (sparse) PCs (Figure~\ref{fig:LymphomanaBiplot1}) visually separates the three types of lymphoma malignancies.

For both PCA results in Figure~\ref{fig:LymphomanaBiplot1}, DLBCL samples are mapped to the left, while FL and CLL observations are mapped to the right and are distinguishable vertically. Thus, the first PC differentiates DLBCL from FL and CLL, and the second PC separates FL from CLL.

\begin{figure*}[h]
\center
\includegraphics[width=.9\textwidth]{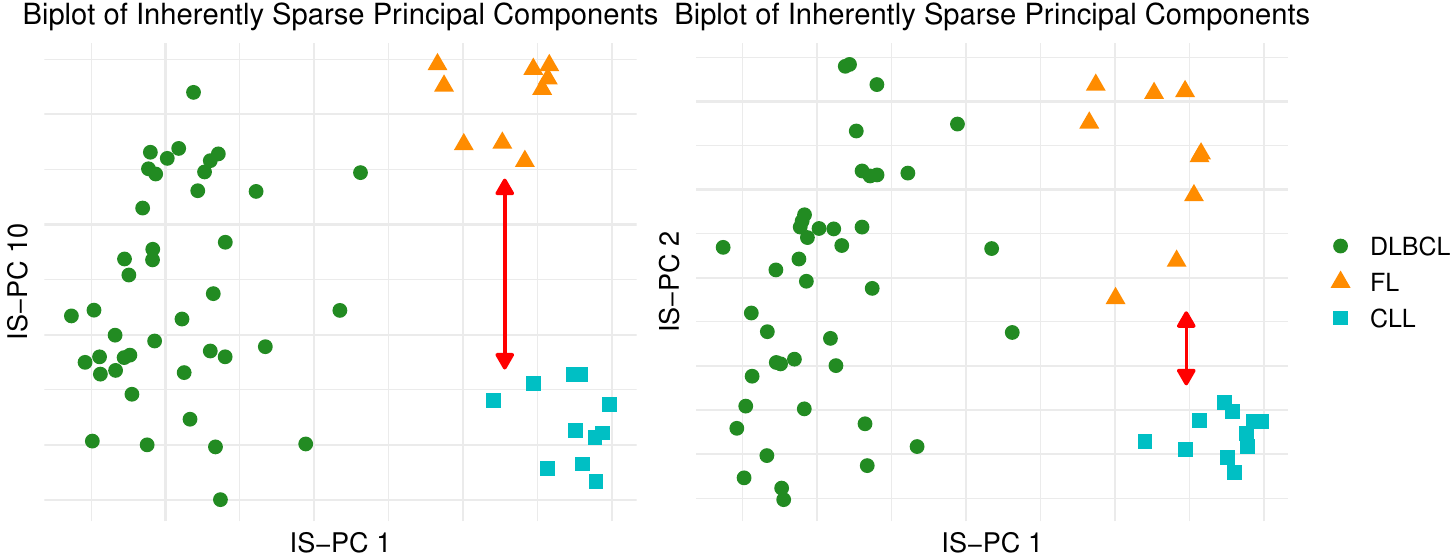}
  \caption{Biplots of the first PCs for the lymphoma gene expression data set, which contains $42$ diffuse large B cell lymphoma (DLBCL) observations \emph{(green, $\bullet$)}, $9$ follicular lymphoma (FL) observations \emph{(orange, $\blacktriangle$)}, and $11$ chronic lymphocytic leukemia (CLL) observations \emph{(blue, $\blacksquare$)}. The left plot shows the first and tenth sparse PC, and the right plot the first and second sparse PC respectively.}
  \label{fig:LymphomanaBiplot2}
\end{figure*}

Notably, the tenth sparse singular vector of IS-PCA improves the separation between FL and CLL by increasing the distance between these two groups. This is shown in Figure~\ref{fig:LymphomanaBiplot2} \emph{(left)}, in comparison to the second sparse singular vector \emph{(right)}. The tenth sparse singular vector contains $22$ nonzero components, seemingly containing mostly genes that provide differences between FL and CLL. It is worth noting that the tenth sparse singular vector is weakly correlated with all non-sparse singular vectors, with the largest correlation being $0.25$ with the second singular vector. This suggests that the $22$ genes in the submatrix underlying the tenth sparse singular vector contain information from multiple non-sparse singular vectors to distinguish FL and CLL. While the second non-sparse singular vector is visualized in Figure~\ref{fig:LymphomanaBiplot1}, visualizations for the third, fourth, and fifth singular vectors are provided in the online appendix, demonstrating that none of them separates FL and CLL. Additionally, we provide code to visualize the non-sparse PCs up to order 16 in the supplementary data.

\section{Discussion}\label{s:Discussion}

In this work, we introduce a methodology for computing sparse PCs in high-dimensional settings that reflect the underlying sparse structure of the data matrix. The proposed approach offers three main advantages over existing methods. First, there is no PCA related method that identifies the sparse covariance structure of data sets. Second, the proposed sparse PCs mirror the inherent structure of the data matrix, avoiding the enforcement of sparsity that may not align with the true data structure. This is desirable, as we have shown that computing sparse singular vectors misaligned with the underlying data structure leads to poor results. Third, the sparse PC loadings are (approximately) orthonormal and therefore do not share common information. This allows computation of the explained variance directly using the eigenvalues (singular values), as in standard PCA.

We demonstrate the practicality and relevance of the proposed methodology through real data applications in the context of gene expressions. The sparse singular vectors and the decomposition of the data matrix into uncorrelated and disjoint submatrices enables the identification of variables contributing to specific characteristics of the gene expressions, providing new additions for PCA.

Building upon this work, we believe that the relationship between principal submatrices and principal variables offers a potential direction for future research. Intuitively, variables contained in a submatrices that contribute little to the overall explained variance are unlikely to contribute substantially in a linear sense to the overall variance, as they are uncorrelated with the variables in the other submatrices. While the identification of principal variables is computationally intensive, restricting the search to submatrices that contribute substantially to the overall variance could make this process more efficient.

\section*{Software and replication}
The methodology has been implemented in the \texttt{R} package \texttt{bdsvd} \citep{BH24}. The data used in the real data examples in Section~\ref{s:RealData} were obtained online at \href{https://www.stat.cmu.edu/~jiashun/Research/software/}{https://www.stat.cmu.edu/$\sim$jiashun/Research/software/}. The prostate tumor data set originates from \citet{SI02} and was mean-centered for this analysis. The lymphoma malignancies data was mean-centered and $\log_2$ transformed, as described in \citet{DFS02}. Additional information about this data can be found in \citet{AL00}.

\section*{Supplementary data}
Supplementary data are available are available upon request.

\nocite{KKC12}
\nocite{SS90}
\nocite{WLL09}
\nocite{WT23}
\nocite{YA09}

\bibliography{sn-bibliography.bib}

\end{document}